\documentclass[aps,prl,twocolumn,superscriptaddress]{revtex4-1}
\usepackage{amsmath}
\usepackage{color}
\usepackage{graphicx}
\usepackage{epstopdf}
\usepackage{color}
\begin{document}

\title{Measurement of a wavelength sized optical vortex knot}%}%
\author{I. A. Herrera-Hern\'andez}
\author{C. A. Mojica-Casique}
\author{P. A. Quinto-Su}%
 \email{pedro.quinto@nucleares.unam.mx}
\affiliation{%
 Instituto de Ciencias Nucleares, Universidad Nacional Aut\'onoma de M\'exico, Apartado Postal 70-543, 04510, Cd. Mx., M\'exico}

%% To be edited by editor
% \dates{Compiled \today}
%\ociscodes{(140.3490) Lasers, distributed feedback; (060.2420) Fibers, polarization-maintaining;(060.3735) Fiber Bragg gratings.}
%\ociscodes{(350.4855)  Optical tweezers or optical manipulation; (140.7010)   Laser trapping; (090.1995)   Digital holography;  (060.4250)   Networks. } 
%% To be edited by editor
% \doi{\url{http://dx.doi.org/10.1364/XX.XX.XXXXXX}}
%\maketitle
%\linenumbers

\begin{abstract}
In the past decade, optical vortex knots with axial dimensions on the order of $10^2-10^5$ wavelengths ($\lambda$) have been implemented in the laboratory. 
However, many potential applications require a drastic reduction on the the size of these knots to the order of $\lambda$.
In this work, we show the first measurement of an optical vortex knot  (trefoil) with an axial length  of $1.99\, \lambda$ contained within a volume ($x\times y\times z$) of $1.90\times 2.19\times 1.99 \, \lambda^3$. Prior to focusing, the laser light is linearly polarized ($x$, horizontal) and we observe the knot in the dominant $x$ polarization component of the tightly focused field. We also observe the transition of the trefoil knot to a couple of vortex loops via reconnection of the vortex lines as we change the spatial scale of the angular spectrum representation of the field.
\end{abstract}
%%%%%%%%%%%%%%%%%%%%%%%%%%%%%%%%%%%%%%%%%%%
\maketitle

Knotted structures have been part of physical models since 1869 \cite{kelvin} and are solutions to several field theories \cite{knotfields}.
Only recently it has been possible to generate fields exhibiting isolated knots in the laboratory, but most of those experimental realizations are short lived and all have macroscopic spatial scales on the range between a few hundreds of micrometers and several centimeters.
For example, in liquid, vortex knots \cite{knotwater} have been created by the sudden immersion of a hydrofoil that has the desired shape (trefoil) with a size of a few centimeters. The knots appear for a few milliseconds and latter (due to viscosity) the vortex lines reconnect and then separate into vortex loops before disappearing. 
Transient knots have also been observed in Bose-Einstein condensates \cite{knotbec} at spatial scales of a few hundreds of micrometers in very brief time intervals on the order of 500 microseconds.

Stable optical knotted structures have been created with polarization \cite{knotcirc, knotpolarizacion, knotinfo} and phase singularities \cite{dennis} in the paraxial regime. 
In particular, optical vortex knots can be generated with 2 dimensional phase masks. The first experimental implementation \cite{dennis} of optical vortex knots generated these structures with axial lengths of a few centimeters.
A similar approach  was used to generate acoustic vortex knots \cite{acusticknots} with an axial length of about 40\,cm.

There is great interest in miniaturizing optical vortex knots to spatial scales on the order of a wavelength ($\lambda$) to enable applications in matter control \cite{halina2017}, information \cite{knotinfo} and microfabrication \cite{microfab}.
So far, the smallest optical vortex knots (called `ultra-small') have axial spatial scales of about 240\,$\lambda$ \cite{ultrasmall} spanning a volume of $10^6\,\mu\mathrm{m}^3 \sim 3.94\times 10^6 \lambda ^3$, 6 orders of magnitude smaller than the previous ones \cite{dennis}. 

Theoretically, it has been shown that optical vortex knots with spatial scales on the order $\lambda$ should be achievable by non paraxial propagation of a tightly focused structured beam \cite{dennisnp, hughes1, hughes2}. 
However, at those spatial scales it is very difficult to get reliable measurements of optical singularities where the light intensity vanishes.  Furthermore, the problem is exacerbated by the fact that (to our knowledge) there are no measurements of tightly focused light fields in small volumes (measurements across multiple spatial transverse planes).

%%%%%%%%%%%%%%%%%%%%%%fig1

\begin{figure}
\includegraphics[width=3.1in]{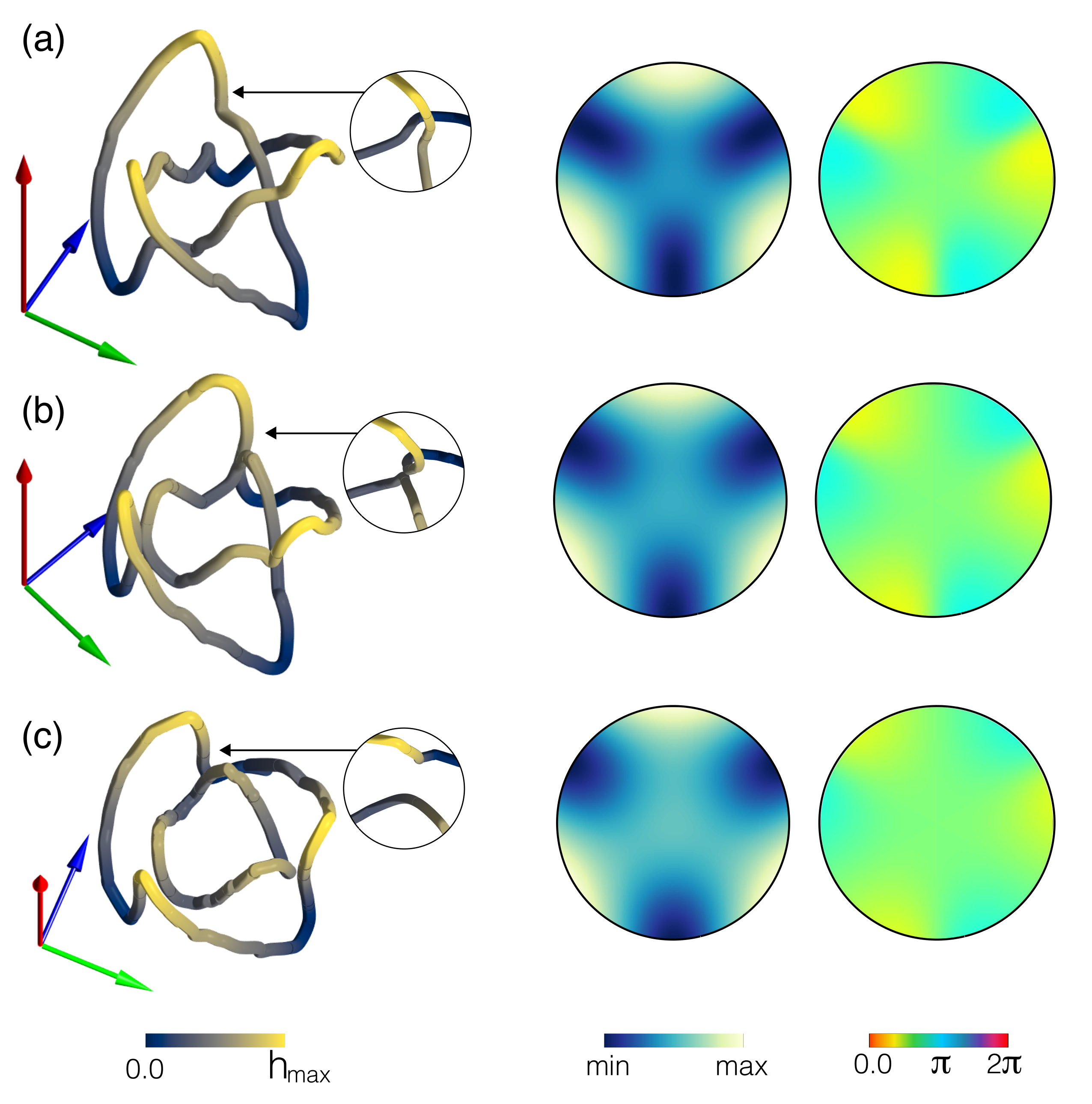} %im11.png 
\caption{Simulated optical vortex lines generated (first column) with the angular spectrum of eq. (1) (last two columns: amplitude and phase). (a) trefoil vortex knot ($w_0=0.75$, $S=0.8$) spanning a volume of $2.07 \lambda \times 2.05 \lambda \times 2.11 \lambda$. (b) Reconnection ($w_0=0.75$, $S=1.0$) contained in $2.04 \lambda \times 2.02 \lambda \times 1.75 \lambda$. (c) Separated loops ($w_0=0.75$, $S=1.2$) spanning $2.06 \lambda \times 2.09 \lambda \times 1.17 \lambda$. The insets in the first column show a view from the top an specific region to illustrate the change in topology. Axes: $x$ axis: green, $y$ axis: blue, $z$ axis: red.}
\end{figure}

%%%%%%%%%%%%%%%%%%%%%%%%%%%%

%%%%%%%%%%%%%%%%%%
Here, we use the angular spectrum calculated in \cite{dennisnp} to create a wavelength sized optical vortex knot, combined with a classical interferometric method to measure tightly focused fields with no approximations \cite{measurefield}. 
We show that it is possible to create a trefoil optical vortex knot enclosed by a volume ($x\times y\times z$) of $1.90\times 2.19 \times 1.99 \,\lambda ^3$. Our work represents an almost 6 orders of magnitude smaller volume compared with the ultra-small knots \cite{ultrasmall} and about 12-13 orders of magnitude with respect to the other paraxial realizations \cite{dennis, knotinfo}.

The angular spectrum representation for a non paraxial trefoil knot is described in \cite{dennisnp} (another approach in \cite{hughes1, hughes2}), where they calculated the Fourier transform of their previous \cite{dennis} paraxial polynomial representation and then selected the terms that can give rise to a knot when propagating the beam in the non paraxial regime with the Richards-Wolf integral \cite{rwolf}.
The complex angular spectrum representation of the non-paraxial trefoil knot in cylindrical coordinates is \cite{dennisnp}:
\begin{equation}
\hat{E}(\rho_k, \varphi_k)=E_0e^{-\rho_k^2/2\omega_0^2}\left(2+3i\rho_k^3 S^{-3}e^{3 i\varphi_k} +3\rho_k^4S^{-4}\right)
\end{equation}
where $\rho_k=(\mathrm{NA}k_0)^{-1}\sqrt{k_x^2+k_y^2}$ is the dimensionless radial coordinate, $k_0$ is the wave number in vacuum and NA the numerical aperture of the system, with this definition $\rho_k=1.0$ at the aperture, $\varphi_k=\tan^{-1}\left(k_y/k_x \right)$, $E_0$ is the amplitude,  $S$ is a scale parameter that determines the knot size and $\omega_0$ is the dimensionless Gaussian waist.

The simulation considers linearly polarized light (horizontal) imprinted with the complex angular spectrum representation (eq. 1) with $w_0 =0.75$ and $S$ in a range between $0.8$ and $1.2$ which is tightly focused by a high numerical aperture lens (NA$=1.11$, refractive index n=1.518). The focused field acquires measurable polarization components in the other  directions ($y$ and $z$). However, the vortex lines that can exhibit a knotted geometry only appear in the $x$ polarization component of the tightly focused field.

Figure 1 shows the calculated optical vortex lines with their angular spectrum (amplitude and phase) for the cases of $S=0.8, \, 1.0, \, 1.2$ (constant $w_0=0.75$). 
The first case (Fig. 1a) with $S=0.8$ results in an isolated trefoil knot with a size ($x\times y\times z$) of $2.07\times 2.05 \times 2.11\, \lambda^3$ in the dominant $x$ component which has about $83.5\%$ of the power ($16.0\%$ in the $z$ component and $0.5\%$ in $y$). 
Changing the scale parameter to $S=1.0$ (Fig. 1b) results in a re-connection of vortex lines at one vertex with about $84.4\%$ of the power ($15.1\%$ in the $z$ component and $0.5\%$ in $y$) with a size of $2.04\times 2.02 \times 1.75 \lambda^3$.  
A further increase in the value of $S$ to $1.2$ (Fig. 1c) results in breaking the knot into two separated loops that span a volume of $2.06\times 2.09 \times 1.17 \lambda^3$ in the $x$ component with about $85.4\%$ of the power ($14.2\%$ in the $z$ component and $0.4\%$ in $y$). 
Notice that the transition from a knot to separated loops when increasing $S$ is similar to what has been observed in transient systems \cite{knotwater, pre}. 

%%%%%%%%%%%%%%%%%%%%%%%%%%%%%%%%%%%%%%%%%%%%%%%%%%%%%%%%%%%%%%

%%%%%%%%%%%%%%%%%%%%%%  fig2
\begin{figure}
\includegraphics[width=2.8in]{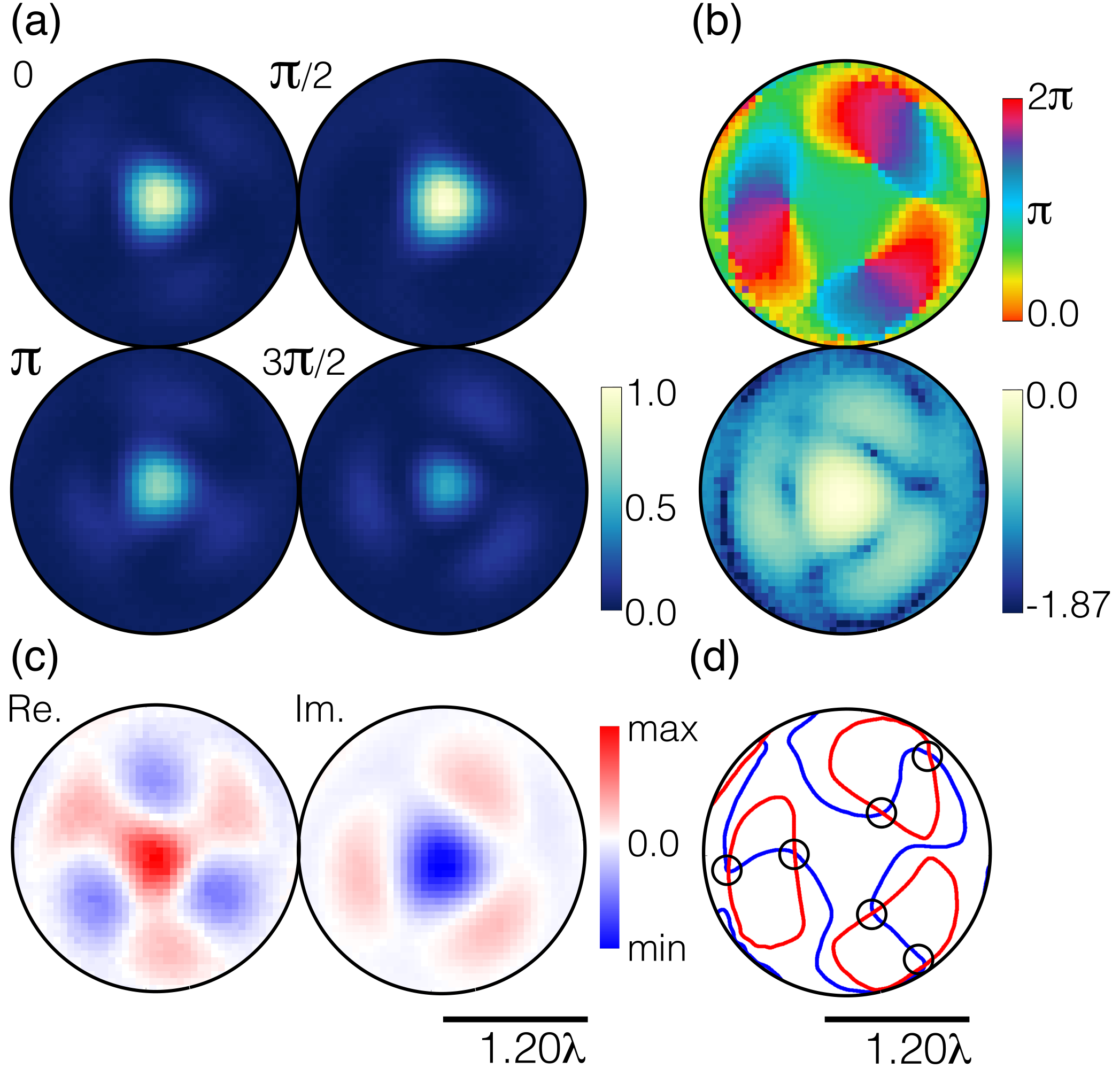}  %3in 
\caption{Measured interferograms to reconstruct $E_x \exp{(i\phi _x)}$ in a single plane ($S=0.8$, $z_0= 1.05 \lambda$ measured relative to the lowest height in detected knot). (a) Interferograms ($\Delta \phi _i=0,\,\pi /2,\,\pi ,\,3\pi /2 $). (b) Extracted phase $\phi _x$ and amplitude $E_x$. The normalized amplitude is in logarithmic scale in order enhance the contrast.  Real and Imaginary part of $E_x\exp{(i\phi _x)}$ (a) (normalized). Real part: min$=-0.38$, max$=0.38$, imaginary part: min$=-0.93$, max$=0.93$. (d) Extracted contours for the zero lines in the Real (blue) and imaginary (red) part of the x-component of the field. The intersections are the positions of the singularities. }
\end{figure}
%%%%%%%%%%%%%%%%%%%

%%%%%%%%%%%%%%%%%%%%fig3
\begin{figure*}
\includegraphics[width=6.5in]{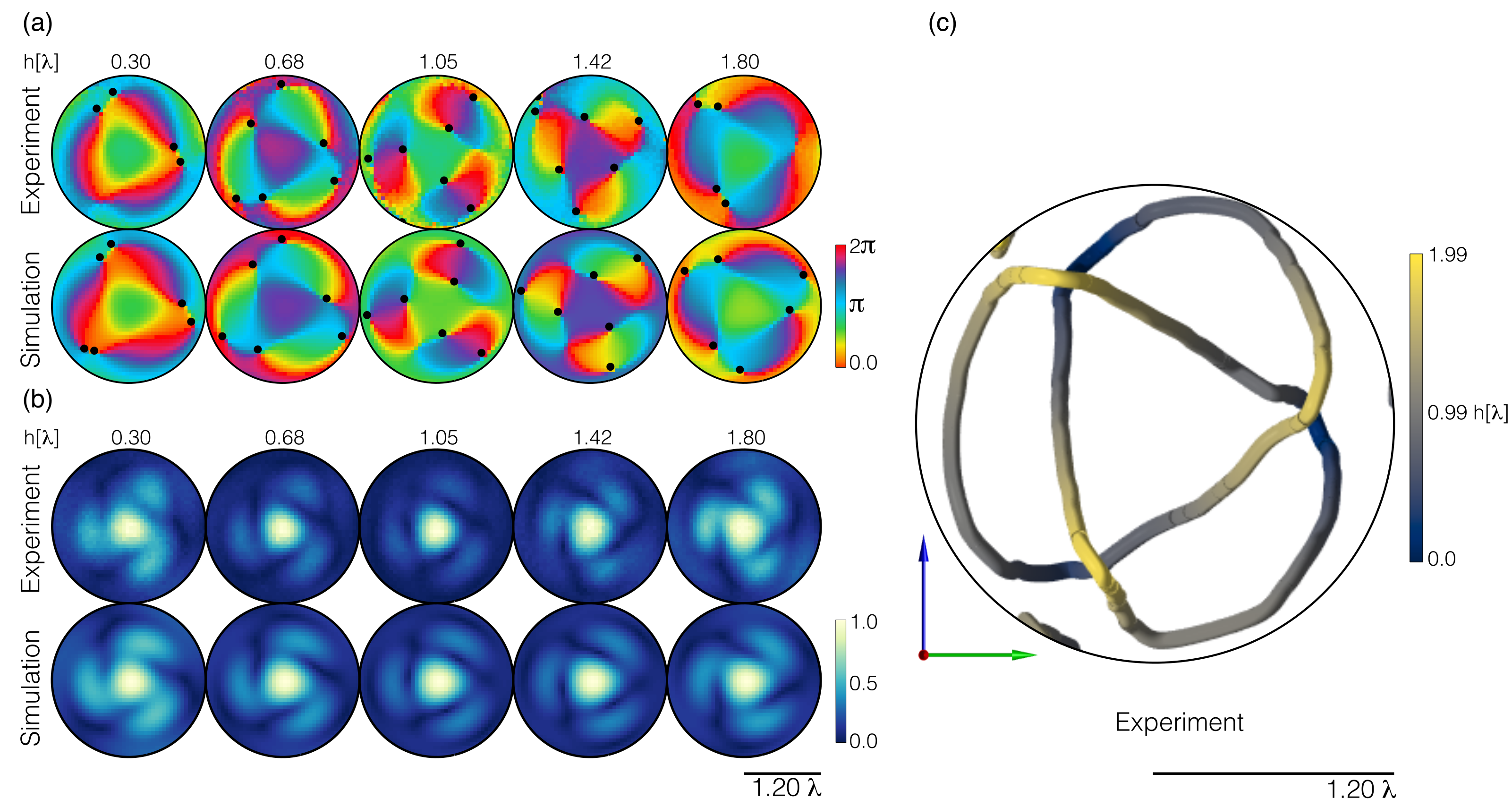} 
\caption{Measuring a trefoil knot. The knot is generated with the parameters $w_0=0.75$ and $S_0=0.8$. (a) Measured and simulated phase $\phi _x$ across multiple planes for the case of  (trefoil knot). The position of the optical vortices are represented by the dots. (b) Measured and simulated amplitudes. Frame diameters: $2.40\, \lambda$. (c) Isolated trefoil knot (top view) reconstructed from the positions of the optical vortices (dots (a)) across 54 planes separated by 40\,nm. The colorbar represents the axial position. The isolated knot is contained in a volume of of $1.90 \lambda \times 2.19 \lambda \times 1.99 \lambda $. Axes: $x$ axis: green, $y$ axis: blue, $z$ axis: red. }
\end{figure*}
%%%%%%%%%%%%%%%%%%%

The experimental setup and the method to measure tightly focused optical fields with no approximations is described in \cite{measurefield} and in the Supplemental Material \cite{supplementalmaterial}. 
The laser light $\lambda =1064\,$nm (Gaussian profile) is shaped with a phase only spatial light modulator (SLM) that generates the beams with the angular spectrum of eq. (1). As the Gaussian beam overfills the SLM, the intensity across the screen is fairly constant which allows us to choose $w_0$ when the phase and amplitude are modulated with the exact complex modulation algorithm of \cite{boyd}. We use the same parameters of the simulations ($w_0=0.75$ and $S=0.8\,, 1.0\,, 1.2$) with an NA of $1.11$ which is controlled by the diameter of the hologram at the SLM.
The field is measured at 80 transverse planes axially separated by 40\,nm (piezo stage). The axial scan starts in a position of $-1.6 \mu$m ($\sim 1.5\,\lambda$) relative to the waist ($z=0$) and increasing until reaching $+1.6\,\mu$m.

The amplitude and phase of the focused field are extracted with 4 phase shift interferometry \cite{measurefield} where we use a diverging reference beam. In this case we only measure the $x$ component that contains the knotted field.  
At a given axial position $z_0$ the measured interferograms are represented by $\mathcal{I}_i(x,y, z_0)$ with $i=1-4$ where a constant phase is added each hologram $\Delta\phi _{1-4}=0, \pi /2 , \pi , 3\pi /2$.

The phase $\phi _x(x,y, z_0)$ and amplitude $E_x(x,y, z_0)$ of the $x$ polarization component of the tightly focused field are directly extracted from the raw data ($\mathcal{I}_i(x,y, z_0)$) with no filters using the relations (removing the spatial dependence to simplify the notation):

\begin{equation}
E'_x=\sqrt{(\mathcal{I}_1-\mathcal{I}_3)^2+(\mathcal{I}_4-\mathcal{I}_2)^2 } 
\end{equation}
and 
\begin{equation}
\phi '_x=\tan^{-1}\left(\frac{\mathcal{I}_4-\mathcal{I}_2}{\mathcal{I}_1-\mathcal{I}_3}\right). 
\end{equation}
where $E'_x \propto E_x E_r$, $\phi '_x=(\phi _x -\phi _r)$, $E_r$, $\phi _r$ are the amplitude and phase of the reference beam. 
Close to the optical axis, both $\phi _r$ and $E_r$ are almost constant due to the large radius of curvature of the diverging reference: $E_r(x,y, z_0)\approx c_1$, $\phi _r(x,y, z_0)\approx c_2$. In this way $E_x (x,y, z_0)\approx E'_x (x,y, z_0)/c_1$ and $\phi _x (x,y, z_0)\approx \phi ' _x (x,y, z_0) + c_2$. To obtain the position of the phase singularities the value of the constants is not important, but when we compare the measured phases to the calculated ones, we choose $c_2$ to minimize a normalized cross correlation \cite{ncc} (NCC). The amplitudes (calculated and measured) are normalized and also compared using the NCC.

Figure 2a shows the 4 interferograms $\mathcal{I}_i$ at $z_0= 1.05 \lambda$ for $S=0.8$, while the reconstructed phase and amplitude are in Fig. 2b. 
The extracted real and imaginary parts of the x-component of the field  Re$(E_x \exp{(i\phi _x)}) \approx \mathcal{I}_1 -\mathcal{I}_3 $ and Im$(E_x\exp{(i\phi _x)}) \approx \mathcal{I}_4 -\mathcal{I}_2 $ are in Fig. 2c. 
Figure 2d depicts the zero contours for the real and imaginary parts of the field and the intersections locate the singularities.

%%%%%%%%%%%%%%%%%%%%%%%%%%%
%%%%%%%%%%%%%%%%%%%% fig4
\begin{figure*}
\includegraphics[width=6.5in]{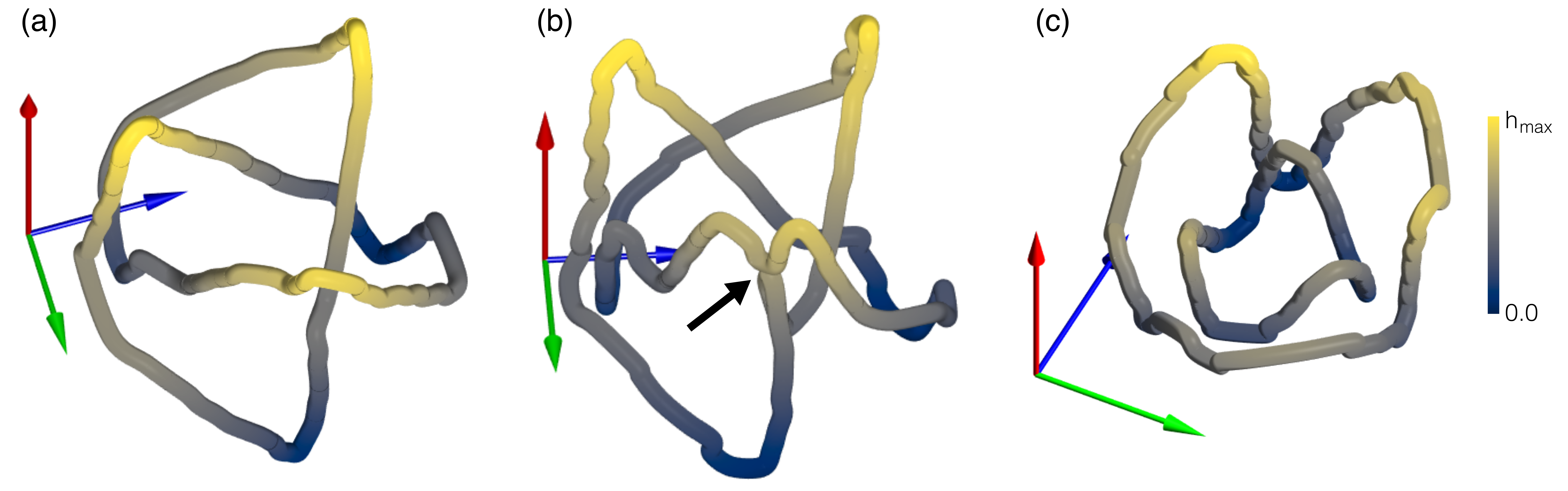} 
\caption{Measured isolated knots and loops varying. The parameter $w_0$ in eq. (1) has a fixed value of 0.75. The colorbar represents the axial position. (a) Trefoil knot ($S=0.8$), size:  $1.90\lambda \times 2.19 \lambda \times 1.99 \lambda $, $h_{max} =1.99\,\lambda$. (b) Reconnection (indicated by the arrow) obtained by increasing the parameter $S$ to $S=1.0$,  size: $1.90\lambda \times 2.13 \lambda \times 1.50 \lambda $, $h_{max} =1.50\,\lambda$. (c) Two separated loops ($S=1.2$), size: $1.89\lambda \times 2.25 \lambda \times 0.75 \lambda $, $h_{max} =0.75\,\lambda$. Axes: $x$ axis: green, $y$ axis: blue, $z$ axis: red.}
\end{figure*}
%%%%%%%%%%%%%%%%%%%

%%%%%%%%%%%%%%%%%%%%%%%%
The extracted phase and amplitude (simulated and measured) of a trefoil knot with $w_0 =0.75$ and $S=0.8$ at different axial heights are in Figure 3a-b. Notice that we labelled the heights $h$ starting from the bottom of the knot. In contrast, the waist of the simulated beam (where the polynomial is defined) is at $z=0$. In the experiment, this plane is located close to the middle of the axial scan (at $h=1.05 \lambda$ in Fig. 3a). More details about the comparison are in the Supplemental Material \cite{supplementalmaterial}.

Figure 3c shows a view from the top of the optical vortices that make the isolated trefoil knot and some other surrounding vortex lines that are not connected to the knot. The colorbar represents the height with respect to the bottom of the knot. 
The isolated knot is contained in a volume of $1.90\times 2.19 \times 1.99 \lambda ^3$, which is smaller than the calculated one. The slight difference in dimensions might be caused by minor errors in the experiment (aberrations and in the implementation of the complex angular spectrum).

Nevertheless, there is good agreement between measurements and simulations with NCC mean values of $0.90 \pm 0.03$ for the phase and $0.97 \pm 0.01$ for the amplitude. 
The isolated knot is bounded on the top and bottom by a phase that has a triangular shape. In the experiment, those bounds (Supplemental Material \cite{supplementalmaterial}) are reached in a shorter axial range ($1.99\,\lambda$).

Figure 4 shows the measured isolated knots (and loops) as a function of the scale parameter $S$. Figure 4a depicts the trefoil knot (Fig. 3) in 3D with $S=0.8$, while Fig. 4b shows the knot generated with $S=1.0$ with a reconnection (indicated by an arrow). We observe that the the reconnection appears at a different vertex than in the simulation. Further increasing $S$ to 1.2 results in two separated loops (Fig. 4c). All cases are contained in a similar transverse area ($\sim 4-4.3 \lambda^2$) and have axial dimensions of 1.99, 1.50, 0.75\,$\lambda$ for $S=0.8,\, 1.0,\, 1.2$ respectively.
The measured vortex lines (knots and loops) are slightly smaller than the calculated ones (Fig. 1) which also span transverse areas with the same range of $ 4-4.3 \lambda ^2$ and axial sizes of $z=2.11,\, 1.75, \, 1.17 \,\lambda$ for $S=0.8,\, 1.0,\, 1.2$ respectively.
Interestingly, the axial dimensions of the measured and simulated knots and loops are similar to the transverse sizes as calculated in \cite{dennisnp}, in contrast with the paraxial realizations \cite{dennis, knotinfo}.

The normalized cross correlations for $S=1,\,1.2$ have values of $(0.89\pm 0.1)$/$(0.97\pm 0.01)$ and $(0.92\pm 0.02)$/$(0.97\pm 0.01)$ respectively (phase/amplitude).
The comparisons between measurements and simulations for $S=1, \,1.2$ are shown in the Supplemental Material \cite{supplementalmaterial}).

Future work will involve other polarization states (e.g. circular) and exploring other possible angular spectrum representations \cite{hughes1, hughes2}.

%%%%%%%%%%%%%%%%%%%%%%%%%%%%%%%%%%%
\subsection*{Acknowledgements}
Work partially funded by DGAPA UNAM PAPIIT grant IN107719, CTIC-LANMAC 2021 and CONACYT LN-299057. IAHH thanks PAPIIT for a scholarship.  
Thanks to Jos\'e Rangel Guti\'errez for the fabrication of some optomechanical components. 

\bibliographystyle{apsrev4-1}
%\bibliographystyle{}

%%%%%%%%%%%%%%%%%%%%%%%%%%%%%%%%%%%%%%%%%%%%%%%%%

\end{document}